\documentclass[pre,twocolumn,notitlepage,tightenlines,%
showpacs,showkeys]{revtex4}

\usepackage{amsmath}
\usepackage{amsfonts}
\usepackage{bm}
\usepackage{euscript}
\usepackage{graphicx}
\usepackage{wasysym}
\usepackage{psfrag}

\graphicspath{{figs/}}

\newcommand{\mathnotation}[2]{\newcommand{#1}{\ensuremath{#2}}}

\renewcommand{\l}{\left} 
\renewcommand{\r}{\right} 
\mathnotation{\pd}{\partial} 
\mathnotation{\ee}{{\mathrm e}} 
\mathnotation{\imi}{\mathrm{i}} 
\mathnotation{\ldef}{\mathrel{\raisebox{.069ex}{:}\!\!=}}
\mathnotation{\rdef}{\mathrel{=\!\!\raisebox{.069ex}{:}}}
\mathnotation{\dint}{\,{\mathrm{d}}} 

\mathnotation{\grad}{\nabla} 
\mathnotation{\lapl}{\Delta} 
\mathnotation{\sdim}{d} 

\renewcommand{\time}{t} 
\mathnotation{\pdt}{\partial_\time} 
\mathnotation{\xc}{x} 
\mathnotation{\yc}{y} 
\mathnotation{\xv}{{\bm{\xc}}} 
\mathnotation{\yv}{{\bm{\yc}}} 
\mathnotation{\kc}{k} 
\mathnotation{\kv}{{\bm{\kc}}} 
\mathnotation{\uc}{u} 
\mathnotation{\uv}{\bm{\uc}} 

\mathnotation{\Diff}{\kappa} 
\mathnotation{\ssc}{s} 
\mathnotation{\lavg}{\langle} 
\mathnotation{\ravg}{\rangle} 
\mathnotation{\Pe}{\mathrm{Pe}} 
\mathnotation{\ME}{\mathcal{E}} 

\mathnotation{\gradpow}{p}
\mathnotation{\xx}{\kc} 
\mathnotation{\lb}{a} 

\begin{document}

\title{Multiscale Mixing Efficiencies for Steady Sources}
\author{Charles R. Doering}
\email{doering@umich.edu}
\affiliation{Department of Mathematics and Michigan Center for Theoretical
Physics, University of Michigan, Ann Arbor, MI 48109-1043, USA}
\author{Jean-Luc Thiffeault}
\email{jeanluc@imperial.ac.uk}
\affiliation{Department of Mathematics, Imperial College
  London, SW7 2AZ, United Kingdom}


\begin{abstract}
Multiscale mixing efficiencies for passive scalar advection are defined in
terms of the suppression of variance weighted at various length scales.  We
consider scalars maintained by temporally steady but spatially inhomogeneous
sources, stirred by statistically homogeneous and isotropic incompressible
flows including fully developed turbulence.  The mixing efficiencies are
rigorously bounded in terms of the P\'eclet number and specific quantitative
features of the source.  Scaling exponents for the bounds at high P\'eclet
number depend on the spectrum of length scales in the source, indicating that
molecular diffusion plays a more important quantitative role than that implied
by classical eddy diffusion theories.
\end{abstract}

\maketitle

\emph{Introduction.}  Mixing in fluid flows plays a central role in many
scientific and engineering applications and is the subject of a large body of
theoretical research~\cite{ADreviews}.  In this paper we consider the mixing
of a passive scalar maintained by a temporally steady but spatially
inhomogeneous source {\it without} a clear separation of length scales between
the source and the stirrer, as is necessary for multiscale analyses, or when
length scales in the source are the smallest scales in the problem.

The effectiveness of a given stirring velocity vector field in a statistically
stationary state is naturally gauged by its ability to suppress the space-time
averaged variance of the scalar: a decrease in the variance indicates a more
uniformly mixed scalar.  The ratio of the scalar variance in the presence of
stirring to that resulting from molecular diffusion alone defines a
dimensionless measure of the mixing efficiency of the flow.  Scalar variances
may be weighted at diverse spatial length scales, so we introduce a family of
mixing efficiencies that measure the effectiveness of the stirring on
different scales.  Not unexpectedly, these mixing efficiencies generally
depend on details of both the source and the stirring.  Sawford \&
Hunt~\cite{Sawford1986} highlighted the role of source size in particle
dispersion models for the case of small sources, showing that the
concentration variance depends explictly on the molecular diffusivity.  Here
we make direct use of the advection--diffusion equation to show the nature of
this dependence for a wide range of source types.

For statistically stationary, homogeneous and isotropic flows---including
turbulence---we derive new, mathematically rigorous and physically relevant
limits on these efficiencies in terms of the P\'eclet number and quantitative
features of the source alone.  We discover that the high-P\'eclet number
scaling exponents of the efficiency bounds depend on the spatial dimension,
the length scales upon which the variance is observed, and spatial scales in
the source.  These results show that the spectrum of length scales in the
source may be more important than the distribution of length scales in the
flow in determining the mixing efficiencies.  Moreover, we derive novel
quantitative estimates of nontrivial scaling exponents for the mixing
efficiencies in cases where the source is fractal or measure-valued.  These
are anomalous scalings that cannot be deduced from dimensional arguments
alone.

\emph{Problem statement and definitions.}  The advection--diffusion equation
for the concentration~$\theta(\xv,\time)$ of a passive scalar is
\begin{equation}
  \pdt\theta + \uv\cdot\grad\theta =
  \Diff\,\lapl\theta + \ssc\,,
  \label{eq:AD}
\end{equation}
where~$\Diff$ is the molecular diffusivity.  We restrict attention to
spatially periodic boundary conditions for $\xv \in [0,L]^{\sdim}$.  The
time-independent source~$\ssc(\xv)$ in~\eqref{eq:AD} is taken (without loss of
generality) to have spatial mean zero so that eventually the
concentration~$\theta(\xv,\time)$ will have spatial mean zero as well.

The velocity field $\uv(\xv,\time)$ is given.
It could be the solution of some
dynamical equations or a particular stochastic process, but in any case we
consider it to be a prescribed divergence-free vector field with the following
equal-time single-point statistical properties shared by homogeneous
isotropic turbulence:
\begin{equation}
\begin{split}
\overline{\uc_{i}(\xv,\cdot)} &= 0,
\quad
\overline{\uc_{i}(\xv,\cdot) \uc_{j}(\xv,\cdot)} = \frac{U^{2}}{d} \delta_{ij}
\\
\overline{\uc_{i}(\xv,\cdot)
  \frac{\partial \uc_{j}(\xv,\cdot)}{\partial \xc_{k}}} &= 0,
\quad
\overline{\frac{\partial \uc_{i}(\xv,\cdot)}{\partial \xc_{k}}
\frac{\partial \uc_{j}(\xv,\cdot)}{\partial \xc_{k}}} =
\frac{\Omega^{2}}{d} \delta_{ij}
\end{split}
\label{eq:shif}
\end{equation}
where overbar represents the long-time average (assumed to exist) at each
point in space.  The r.m.s. velocity $U$ measures the strength of the stirring
and~$\Omega$ indicates the flow field's strain or shear content.  The
ratio~$\lambda = U/\Omega$ corresponds to the Taylor microscale for
homogeneous isotropic turbulence.  The P\'eclet number for the flow is $\Pe =
UL/\Diff$.

We quantify the mixing of the scalar by the magnitude of the variances
$\l\lavg\lvert\grad\theta\rvert^2\r\ravg$,~$\l\lavg\theta^2\r\ravg$,
and~$\l\lavg\lvert\grad^{-1}\theta\rvert^2\r\ravg$ where $\lavg\cdot\ravg$
denotes space-time averaging.
(The operator~$\grad^{-1}$ is defined by its Fourier representation in the
periodic domain,~$\grad^{-1} \rightarrow -\imi\kv/\kc^2$.)
These variances measure the fluctuations of~$\theta$ at relatively small,
intermediate and large length scales.
Collectively we write the norms
$\l\lavg\lvert\grad^\gradpow\theta\rvert^2\r\ravg$
for $\gradpow=1$, $0$, and~$-1$.
Note that~$\l\lavg\lvert\grad^{-1}\theta\rvert^2\r\ravg$ is reminiscent of
the recently--introduced ``mix-norm'' \cite{Mathew2005}, as both downplay
the importance of the small scales.

In order to define dimensionless mixing efficiencies we use the baseline
variances defined by the solution $\theta_{0}$ of equation~\eqref{eq:AD} with
the same source but~$\uv=0$, i.e.,~\hbox{$\theta_{0} =
-\Diff^{-1}\lapl^{-1}\ssc$}.  Comparing fluctuations in the presence of
stirring to the moments~$\l\lavg\lvert\grad^\gradpow\theta_{0}\rvert^2\r\ravg$
allows us to gauge the effect of stirring.  We define the dimensionless
\emph{mixing efficiencies} $\ME_\gradpow$ by
\begin{equation}
  \ME_\gradpow^2 \ldef
  \bigl\lavg\lvert\grad^\gradpow\theta_{0}\rvert^2\bigr\ravg/
  \l\lavg\lvert\grad^\gradpow\theta\rvert^2\r\ravg.
  \label{eq:ME}
\end{equation}
These efficiencies increase when the stirring decreases the scalar variances
relative to molecular diffusion alone.

\emph{Mixing efficiency bounds.}  Upper bounds on $\ME_\gradpow$ result
from lower limits on the variances.  We begin with estimates
on~$\l\lavg\theta^2\r\ravg$ using the method developed
in~\cite{Thiffeault2004}: multiply~\eqref{eq:AD} by an arbitrary (but smooth,
spatially periodic) function~$\varphi(\xv)$, average, and integrate by parts
to find
\begin{equation}
  \l\lavg\theta\,(\uv\cdot\grad + \Diff\lapl)\varphi\r\ravg
  = -\l\lavg\varphi\ssc\r\ravg.
  \label{eq:projeq}
\end{equation}
Fluctuations are bounded from below via the min-max variational expression
\begin{equation*}
  \l\lavg\theta^2\r\ravg \ge \max_{\varphi}\min_{\vartheta}
  \l\{\bigl\lavg\vartheta^2\bigr\ravg
  \,\bigl\lvert\,\,
  \bigl\lavg\vartheta\,(\uv\cdot\grad\varphi
  + \Diff\,\lapl\varphi)\bigr\ravg
  = -\l\lavg\varphi\ssc\r\ravg\r\}.
\end{equation*}
The minimization over $\vartheta$ is easily achieved, equivalent to
application of the Cauchy--Schwarz inequality, yielding
\begin{equation}
  \l\lavg\theta^2\r\ravg \ge \max_{\varphi} 
  \l\lavg\varphi\ssc\r\ravg^2/
  \l\lavg\l(\uv\cdot\grad\varphi + \Diff\lapl\varphi\r)\!{}^2\r\ravg \,.
     \label{eq:Varbound}
\end{equation}
This is the sort of variational estimate derived in~\cite{Thiffeault2004}.
Plasting and Young recently enhanced that analysis by including the scalar
dissipation rate as a constraint~\cite{Plasting2006}.

The subsequent maximization over~$\varphi$ is particularly simple for
statistically homogeneous and isotropic flows satisfying~\eqref{eq:shif}, for
then the denominator in~\eqref{eq:Varbound} is
\begin{equation}
      \l\lavg \l(\uv\cdot\grad\varphi + \Diff\lapl\varphi\r)\!{}^2 \r\ravg
      = \l\lavg \frac{U^{2}}{d}|\grad\varphi|^{2} +
      \Diff^{2}\l(\lapl\varphi\r)^2 \r\ravg,
     \label{eq:den1}
\end{equation}
i.e., the quadratic form $\l<\varphi [\Diff^2\lapl^2 -
(U^2/\sdim)\lapl]\varphi \r>$.  Hence the variational
problem~\eqref{eq:Varbound} yields
\begin{equation}
  \ME_0^2 \le \frac{\l\lavg\ssc\,\lapl^{-2}\ssc\r\ravg}
       {\bigl\lavg\ssc\l\{\lapl^2 - (U^2/\Diff^{2}\sdim)\,
\lapl\r\}^{-1}\ssc\bigr\ravg}\,,
  \label{eq:MEHIT}
\end{equation}
a bound that depends on the ``shape'' of the source function but not its
amplitude, and on the stirring velocity field only through its influence on
the length scale $\Diff/U = \Pe^{-1}L$.

Limits for the small scale and large scale efficiencies $\ME_{\pm 1}$ are
obtained from~\eqref{eq:projeq} in the same manner after
integrations by parts and application of the Cauchy--Schwarz inequality.  For
the gradient variance
\begin{equation*}
  \l\lavg\varphi\ssc\r\ravg^{2} =
  \l\lavg (\uv\varphi + \Diff\grad\varphi)\cdot \grad\theta \r\ravg^{2}
  \le \l\lavg |\uv\varphi + \Diff\grad\varphi|^{2}\r\ravg
  \l\lavg |\grad\theta|^{2} \r\ravg
\end{equation*}
so
\begin{equation}
  \l\lavg\lvert\grad\theta\rvert^2\r\ravg \ge \max_{\varphi}
  \l\lavg\varphi\ssc\r\ravg^2/
  \l\lavg\l(\uv\varphi + \Diff\grad\varphi\r)\!{}^2\r\ravg\,.
  \label{eq:gradphiest}
\end{equation}
A potentially sharper bound involving the full two-point correlation function
for the velocity field can be obtained by formally minimizing over
$\theta$~\cite{STD2006}, but for our purposes the
estimate~\eqref{eq:gradphiest} suffices.  For statistically homogeneous
isotropic flows the denominator above is $\l<\varphi [-\Diff^2\lapl + U^2]
\varphi \r>$ and optimization over $\varphi$ leads to
\begin{equation}
  \ME_1^2 \le \frac{\l\lavg\ssc\,(-\lapl^{-1})\ssc\r\ravg}
     {\l\lavg\ssc\l\{-\lapl + U^2/\Diff^{2}\, \r\}^{-1}\ssc\r\ravg}\,.
  \label{eq:ME1HIT}
\end{equation}

The bound on large scale fluctuations follows from~\eqref{eq:projeq}
using~$\theta=\grad\cdot\grad^{-1}\theta$, an integration by parts, and
Cauchy--Schwarz:
\begin{multline*}
  \l\lavg\varphi\ssc\r\ravg^{2} =
  \l\lavg\grad(\uv\cdot\grad\varphi +
  \Diff\,\lapl\varphi)\cdot(\grad^{-1}\theta)\,\r\ravg^{2} \\
  \le \l\lavg |\grad\uv\cdot\grad\varphi +
  \uv\cdot\grad\grad\varphi +\Diff\lapl\grad\varphi|^{2}\r\ravg
  \l\lavg |\grad^{-1}\theta|^{2} \r\ravg
\end{multline*}
so that
\begin{equation*}
  \l\lavg\lvert\grad^{-1}\theta\rvert^2\r\ravg \ge \max_{\varphi}
  \frac{\l\lavg\varphi\ssc\r\ravg^2}
       {\l\lavg |\grad\uv\cdot\grad\varphi +
	 \uv\cdot\grad\grad\varphi +\Diff\lapl\grad\varphi|^2\r\ravg}\,.
\end{equation*}
For statistically homogeneous isotropic flows the denominator is $\l<\varphi
[-\Diff^2\lapl^3 + (U^2/\sdim)\,\lapl^2 - (\Omega^{2}/\sdim)\, \lapl] \varphi
\r>$ so that
\begin{equation*}
  \ME_{-1}^2 \le \frac{\l\lavg\ssc\,(-\lapl^{-3})\ssc\r\ravg}
       {\l\lavg\ssc\l\{-\lapl^{3} + (U^2/\Diff^{2}\sdim)\,
\lapl^{2} - (\Omega^2/\Diff^{2}\sdim)\,\lapl \r\}^{-1}\ssc\r\ravg}\,.
\end{equation*}

It is helpful to rewrite the bounds in Fourier space:
\begin{subequations}
\begin{gather}
 \ME_{1}^2  \le \frac {\sum_\kv
   \lvert\ssc_\kv\rvert^2/\kc^2}
  {\sum_\kv
  \lvert\ssc_\kv\rvert^2/(\kc^2 + \Pe^2)}\,,
  \label{eq:MEHITb}
\\
  \ME_0^2 \le \frac {\sum_\kv
  \lvert\ssc_\kv\rvert^2/\kc^{4}}
  {\sum_\kv \lvert\ssc_\kv\rvert^2/ (\kc^4 + \Pe^2\kc^2/\sdim)}\,,
  \label{eq:MEHITm}
\\
  \ME_{-1}^2 \le \frac {\sum_\kv \lvert\ssc_\kv\rvert^2/\kc^{6}}
  {\sum_\kv \lvert\ssc_\kv\rvert^2/
  (\kc^6 + \Pe^2\kc^4/\sdim + \Pe^2\kc^2/\lambda^{2}\sdim)}\,,
  \label{eq:MEHITp}
\end{gather}
\end{subequations}
where we have rescaled $[0,L]^{\sdim}$ to $[0,1]^{\sdim}$ so that wavevector
components are integer multiples of $2\pi$.  Now
we investigate the large P\'eclet number behavior of these bounds for a
variety of classes of sources.

\emph{Monochromatic Sources.} For sources that depend only on
a single wavenumber~$\kc_{s}$, the bounds simplify to
\begin{gather*}
  \ME_{1} \le \sqrt{1 + \Pe^2/\kc_{s}^{2}}\,, \\
  \ME_{0} \le \sqrt{1 + \Pe^2/\kc_{s}^{2}\sdim}\,, \\
  \ME_{-1} \le \sqrt{1 + \Pe^2/\kc_{s}^{2}\sdim + 
  \Pe^2/\lambda^{2}\kc_{s}^{4}\sdim}\,.
\end{gather*}
Each efficiency is asymptotically proportional to $\Pe$, corresponding to the
expected suppression of variance if the molecular diffusivity $\Diff$ is
replaced by an eddy diffusivity proportional to $UL$.  Moreover these upper
bounds are sharp: they may be realized by uniform flow fields whose direction
varies appropriately in time to satisfy the weak statistical homogeneity and
isotropy conditions used in the analysis~\cite{STD2006,Plasting2006}.
Each estimate also exhibits a decreasing
dependence on the length scale of the source: at high $\Pe$ the estimates for
the small and intermediate scale efficiencies $\ME_{1}$ and $\ME_{0}$ are
$\sim \Pe/\kc_{s}$.
This suggests that an eddy diffusivity might better be
defined as a product of $U$ and a length scale $\sim \kc_{s}^{-1}$
characterizing the {\em source}, rather than some scale characterizing the
stirring.

\emph{Square Integrable Sources and Sinks.} In situations where the Fourier
coefficients are square-summable, the asymptotic $\Pe \rightarrow \infty$
behavior of the mixing efficiency bounds for smooth sources is straightforward:
\begin{subequations}
\begin{gather}
  \ME_{1} \le \Pe\, \sqrt{ 
  \frac{\sum_\kv \lvert\ssc_\kv\rvert^2/\kc^{2}}
  {\sum_\kv {\lvert\ssc_\kv\rvert^2}}}\,,
  \label{eq:smoothMEboundb} \\
  \ME_{0} \le \Pe\, \sqrt{ 
  \frac{\sum_\kv \lvert\ssc_\kv\rvert^2/\kc^{4}}
  {\sdim  \sum_\kv \lvert\ssc_\kv\rvert^2/\kc^{2}}}\,,
  \label{eq:smoothMEboundm} \\
  \ME_{-1} \le \Pe\, \sqrt{ 
  \frac{\sum_\kv \lvert\ssc_\kv\rvert^2/\kc^{6}}
  {\sdim \sum_\kv \lvert\ssc_\kv\rvert^2/
  (\kc^{4}+\kc^{2}/\lambda^{2})}}\,.
    \label{eq:smoothMEboundp}
\end{gather}
\label{eq:smoothMEbound}
\end{subequations}
\noindent
These are the same $\Pe$ scalings as observed for monochromatic sources but
with prefactors involving {\it distinct} characteristic length scales of the
source tailored to the scales where the different efficiencies are tuned.  We
reiterate that the $\Pe^{1}$ scaling of the efficiencies is precisely that
which is expected from the conventional notion of eddy diffusion, at least
with regard to the $U$ and $\Diff$ dependence.  A novel feature of these
rigorous estimates is that the ``mixing lengths'' to be employed in
constructing the effective diffusion from $U$ depend on (i) the {\em source}
structure rather than some characteristic persistence length in the flow, and
(ii) the length scales in the concentration fluctuations stressed by the
different multiscale efficiencies.

\emph{Rough Sources.} The P\'eclet number {\it scaling} may actually change
for ``rough'' sources, i.e., when $\ssc(\xv)$ is not square integrable,
resulting in anomalous behavior for some of the efficiencies.  The roughest
physically meaningful sources are measure-valued sources like
$\delta$-functions with nondecaying Fourier coefficients~$\lvert\ssc_\kv\rvert
= {\cal O}(1)$ as~$\kc\rightarrow\infty$.  Then the sums
in~\eqref{eq:smoothMEboundb} and the denominator of~\eqref{eq:smoothMEboundm}
diverge in $\sdim = 2$ or~$3$ so those scalings are invalid.  In this case the
$\Pe$ dependence of $\ME_{1}$ disappears altogether and {\it all} finite
kinetic energy stirring fields are completely {\it ineffective} at suppressing
small scale fluctuations.

To determine the high-$\Pe$ behavior of $\ME_{0}$ we approximate sums by
integrals.  The denominator of~\eqref{eq:MEHITm} is
\begin{equation}
  \sum_\kv \frac{1}{\kc^4 + (\Pe^2/\sdim)\,\kc^2} \sim
  \int_{2\pi}^\infty
  \frac{\xx^{\sdim-1}\dint\xx}{\xx^4 + \Pe^{2}\xx^2/(4\pi^{2}\sdim)}.
  \label{eq:sum2int}
\end{equation}

For~$\sdim=2$ the integral in~\eqref{eq:sum2int} is
\begin{equation}
  \int_{2\pi}^\infty
  \frac{\xx \dint\xx}{\xx^4 + \Pe^{2}\xx^2/8\pi^{2}}
  \sim
  \frac{\log \Pe}{\Pe^{2}}\,,
\end{equation}
resulting in the asymptotic bound
\begin{equation}
  \ME_0 \lesssim
  {\Pe}/{\sqrt{\log\Pe}}\,,
  \quad \sdim = 2.
  \label{eq:anom2}
\end{equation}
Hence there is at the very least a logarithmic correction to $\ME_0$ as
compared to the square integrable source case.

For~$\sdim=3$ the integral in~\eqref{eq:sum2int} becomes
\begin{equation}
  \int_{2\pi}^\infty
  \frac{\xx^2\dint\xx}{\xx^4 + \Pe^2\,\xx^2/12\pi^{2}}
  \sim \frac{1}{\Pe}
\end{equation}
resulting in an anomalous scaling bound
\begin{equation}
  \ME_0 \lesssim \sqrt{\Pe}\,, \quad \sdim = 3.
  \label{eq:anom3}
\end{equation}
This is a dramatic modification of the classical scaling.

Similar analysis shows that the upper bound on the large scale mixing
efficiency $\ME_{-1} \sim \Pe$ in~(\ref{eq:smoothMEboundp}) {\em persists}
even for these roughest sources.

\begin{figure}
\psfrag{Pe1/2}{$\,\Pe^{1/2}$}
\psfrag{Pe1}{{$\!\!\Pe$}}
\psfrag{Pe}{\raisebox{-.5em}{$\Pe$}}
\psfrag{M0}{\raisebox{.5em}{$\ME_0$}}
\psfrag{l/L = 1e-08}{$\ell/L=10^{-8}$}
\centerline{\includegraphics[width=.9\columnwidth]{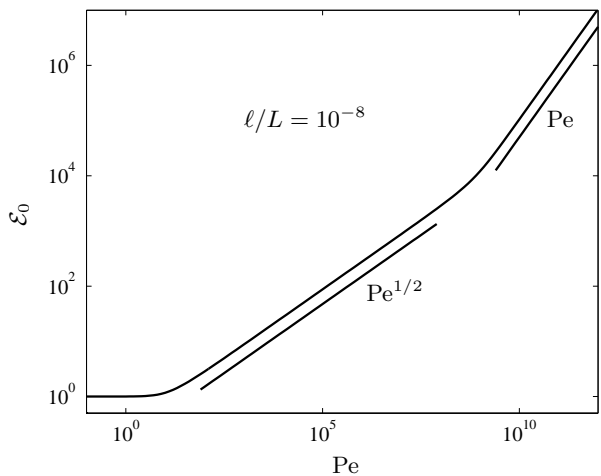}}
\caption{Upper bound for the mixing efficiency $\ME_{0}$ as a function of
P\'eclet number for a small source with $\ell = 10^{-8} L$ stirred by a
three-dimensional statistically homogeneous and isotropic flow [computed from
Eq.~\eqref{eq:MEHITm}].  The intermediate $\Pe^{1/2}$ scaling for~$1 \ll \Pe
\ll (L/\ell)$ is evident.}
\label{fig:mixeff3D}
\end{figure}

\emph{Rough sources with a cutoff.}  Approximate delta-like sources of small
but finite size~$\ell$ have Fourier coefficients~$\ssc_\kv$ that are
approximately constant in magnitude up to a cutoff wavenumber of
order~$2\pi/\ell$, beyond which the spectrum decays as for a smooth source.
We may deduce the behavior of the bound on $\ME_{0}$ for such sources by
inserting an upper limit at~\hbox{$L/\ell \gg 1$} into the integral
in~\eqref{eq:sum2int}.  For large but intermediate P\'eclet numbers
satisfying~$1 \ll \Pe \ll L/\ell$, the cutoff is ineffective so the
logarithmic correction~\eqref{eq:anom2} in $\sdim=2$ and the anomalous
scaling~\eqref{eq:anom3} in $\sdim=3$ appear.  However for $\Pe \gg L/\ell$,
i.e., when the modified P\'eclet number based on the smallest scale in the
source $U\ell/\Diff \gg 1$, the smooth source results apply and we recover the
mixing efficiency bounds linear in~$\Pe$, as in~\eqref{eq:smoothMEbound}.
Figure~\eqref{fig:mixeff3D} shows this scaling transition for the $\sdim=3$
case.  Even in the ultimate regime where the source appears smooth, the
\emph{prefactor} in front of the high-$\Pe$ scaling bounds are significantly
diminished by the small scales in the source: $\ME_0 \lesssim
\l[\log(L/\ell)\r]^{-1/2}\Pe$ in $\sdim = 2$, and $\ME_0 \lesssim
[\ell/L]^{1/2}\Pe$ in $\sdim = 3$.

\emph{Fractal Sources.}  We may also analyze anomalous scalings
for more general ``fractal" rough sources where the Fourier spectrum
$\lvert\ssc_\kv\rvert$ decays as~$\kc^{-\gamma}$ with~\hbox{$0 \le \gamma \le
\sdim/2$}.  The roughest measure--valued sources have~$\gamma=0$ while for
$\gamma > \sdim/2$ the source is square integrable and thus effectively smooth
as far as these multiscale mixing efficiencies are concerned.  In order to
examine the high P\'eclet number asymptotics of the bounds on the various
$\ME_\gradpow$ we estimate integrals similar to~\eqref{eq:sum2int} but with an
extra factor of~$\xx^{-2\gamma}$ in the numerator arising
from~$\lvert\ssc_\kv\rvert^2$.  The results are summarized in
Table~\ref{tab:scalingsall}.  In $\sdim=2$ the scaling for $\ME_{1}$ is
anomalous for any degree of roughness while $\ME_{0}$ is anomalous only
for~$\gamma=0$.  In $\sdim=3$, $\ME_{1}$ is again anomalous for any degree of
roughness while $\ME_{0}$ scales anomalously for~$0\le\gamma<1/2$.  For both
$\sdim=2$ and $3$ the bound on the large scale mixing efficiency $\ME_{-1}$ is
always conventional.  Of course these scalings neglect any large-$\kc$ cutoff
for the rough sources.
If there is a cutoff at wavenumber~$2\pi/\ell$ then the same arguments apply to
recover the normal scaling $\sim \Pe$ for~\hbox{$\Pe \gg L/\ell$}.

\emph{Discussion.}  The multiscale mixing efficiency bounds derived here
reveal new aspects of the ability---and in some cases the {\em inability}---of
statistically homogeneous and isotropic incompressible flows to effectively
suppress fluctuations in passive scalars.  Several notable features of the
phenomena have emerged from this analysis.  One is that the structure of the
source maintaining the scalar concentration plays a central role in efficiency
of mixing at different length scales while the detailed structure of the flow,
i.e., the spectrum of length (and time) scales in the velocity field, plays
only a secondary role as far as the small and intermediate scale efficiencies
$\ME_{1}$ and $\ME_{0}$ are concerned.  The bound on $\ME_{-1}$
in~(\ref{eq:smoothMEboundp}), however, means that small scale structure and
strain in the advecting flow \textit{could} possibly enhance the large scale
mixing efficiency.

For rough sources the efficiencies scale at least as anomalously as indicated
in Table~\ref{tab:scalingsall}, and no clever small scale stirring
can alleviate this effect. 
When efficiency bounds scale anomalously, molecular diffusion plays a much
more important role than implied by conventional eddy diffusion theories.
For then---as far as variance suppression is concerned---there can be then no
``residual'' effective diffusivity due to stirring in the limit of
negligible molecular diffusion.

\begin{table}
\vspace{.5em}
\caption{Scalings of the bound on the mixing efficiency~$\ME_\gradpow$ as
functions of the source roughness exponent~$\gamma$ of the source in two and
three dimensions.}
\begin{flushleft}
\begin{ruledtabular}
\begin{tabular}{lccc}
$\sdim = 2$ & $\gradpow=1$ & $\gradpow=0$ & $\gradpow=-1$ \\
\hline
$\gamma = 0$       & 1 & $\Pe/(\log\Pe)^{1/2}$ & \Pe \\
$0 < \gamma < 1$ & $\Pe^\gamma$ & \Pe & \Pe \\
$\gamma = 1$       & \ \ \ $\Pe/(\log\Pe)^{1/2}$ & \Pe & \Pe \\
$\gamma > 1$       & \Pe & \Pe & \Pe \\[2pt]
\hline
$\sdim = 3$\\ 
\hline
$\gamma = 0$         & 1 & $\Pe^{1/2}$ & \Pe \\
$0 < \gamma < 1/2$   & 1 & $\Pe^{\gamma+1/2}$ & \Pe \\
$\gamma = 1/2$       & 1 & \quad $\Pe/(\log{\Pe})^{1/2}$ & \Pe \\
$1/2 < \gamma < 3/2$ & $\Pe^{\gamma - 1/2}$ & \Pe & \Pe \\
$\gamma = 3/2$       & \ $\Pe/(\log{\Pe})^{1/2}$ & \Pe & \Pe \\
$\gamma > 3/2$       & \Pe & \Pe & \Pe \\
\end{tabular}
\end{ruledtabular}
\end{flushleft}
\label{tab:scalingsall}
\end{table}

\emph{Acknowledgments.}  We are grateful for stimulating discussions with
Francesco Paparella, Tiffany A. Shaw, William R. Young, and many participants
of the 2005 GFD Program at Woods Hole Oceanographic Institution where much of
this work was performed.  CRD. was supported by NSF Grants Nos.\ PHY-0244859
and PHY-0555234 and the Alexander von Humboldt Foundation.  J-LT was supported
in part by the UK Engineering and Physical Sciences Research Council Grant
GR/S72931/01.


\end{document}